\documentclass[twocolumn,showpacs,aps,prl,superscriptaddress,letterpaper]{revtex4}

\usepackage{graphicx}
\usepackage{dcolumn}
\usepackage{amsmath}
\usepackage{epsfig}
\usepackage{multirow}

\long\def\inst#1{\par\nobreak\kern 4pt\nobreak
    {\itshape #1}\par\vskip 10pt plus 3pt minus 3pt}

\RequirePackage{xspace}
\usepackage{relsize}

\def\babar{\mbox{\slshape B\kern-0.1em{\smaller A}\kern-0.1em
    B\kern-0.1em{\smaller A\kern-0.2em R}}}
\def\Abar    {\kern 0.18em\overline{\kern -0.18em A}{}\xspace}
\def\Kbar    {\kern 0.18em\overline{\kern -0.18em K}{}\xspace}
\def\Dbar    {\kern 0.18em\overline{\kern -0.18em D}{}\xspace}
\def\Bbar    {\kern 0.18em\overline{\kern -0.18em B}{}\xspace}

\def\BB      {\ensuremath{B\Bbar}\xspace} 
\def\Bz      {\ensuremath{B^0}\xspace}
\def\Bzb     {\ensuremath{\Bbar^0}\xspace}
\def\BzBzb   {\ensuremath{\Bz {\kern -0.16em \Bzb}}\xspace}
\def\Bu      {\ensuremath{B^+}\xspace}
\def\Bub     {\ensuremath{B^-}\xspace}
\def\Bp      {\ensuremath{\Bu}\xspace}

\def\Bpm     {\ensuremath{B^\pm}\xspace}

\def\BpBm    {\ensuremath{\Bu {\kern -0.16em \Bub}}\xspace}

\newcommand{\optbar}[1]{\shortstack{{\tiny (\rule[.4ex]{1em}{.1mm})}
  \\ [-.7ex] $#1$}}
\def\BorBbar    {\kern 0.18em\optbar{\kern -0.18em B}{}\xspace}
\def\DorDbar    {\kern 0.18em\optbar{\kern -0.18em D}{}\xspace}
\def\KorKbar    {\kern 0.18em\optbar{\kern -0.18em K}{}\xspace}

\def\pep2{PEP-II}
\mathchardef\Upsilon="7107
\def\Y#1S{\ensuremath{\Upsilon{(#1S)}}\xspace}

\def\FourS {\Y4S}

\newcommand{\gevcc}{\ensuremath{{\mathrm{\,Ge\kern -0.1em V\!/}c^2}}\xspace}
\newcommand{\gev}{\ensuremath{\mathrm{\,Ge\kern -0.1em V}}\xspace}
\def\kaon  {\ensuremath{K}\xspace}
\def\Kstarb  {\ensuremath{\Kbar^*}\xspace}
\def\Kp    {\ensuremath{K^+}\xspace}
\def\Km    {\ensuremath{K^-}\xspace}
\def\Kpm   {\ensuremath{K^\pm}\xspace}
\def\Kstar   {\ensuremath{K^*}\xspace}
\newcommand{\etapr}{\ensuremath{\eta^{\prime}}\xspace}
\def\pipi  {\ensuremath{\pi^+\pi^-}\xspace}
\def\KS    {\ensuremath{K^0_{\scriptscriptstyle S}}\xspace}
\def\Kz    {\ensuremath{K^0}\xspace}
\def\Kzb   {\ensuremath{\Kbar^0}\xspace}
\def\pip   {\ensuremath{\pi^+}\xspace}
\def\pim   {\ensuremath{\pi^-}\xspace}
\def\mes        {\mbox{$m_{\rm ES}$}\xspace}
\def\DeltaE     {\mbox{$\Delta E$}\xspace}
\def\epem       {\ensuremath{e^+e^-}\xspace}
\newcommand{\mev}{\ensuremath{\mathrm{\,Me\kern -0.1em V}}\xspace}
\newcommand{\mevcc}{\ensuremath{{\mathrm{\,Me\kern -0.1em V\!/}c^2}}\xspace}

\def\pimp  {\ensuremath{\pi^\mp}\xspace}
\def\BR         {{\ensuremath{\cal B}\xspace}}

\def\beq {\begin{equation}}
\def\eeq {\end{equation}}

\def\etaX {\ensuremath{{\eta_X}}\xspace}
\def\et#1#2#3{\ensuremath{\eta(#1#2#3)}\xspace}
\def\et#1#2#3#4{\ensuremath{\eta(#1#2#3#4)}\xspace}
\def\ph#1#2#3#4{\ensuremath{\phi(#1#2#3#4)}\xspace}
\def\fone#1#2#3#4{\ensuremath{f_1(#1#2#3#4)}\xspace}
\def\x#1#2#3#4{\ensuremath{X(#1#2#3#4)}\xspace}
\def\f_1{\ensuremath{f_1}\xspace}
\def\az{\ensuremath{a_0}\xspace}

\newcommand{\bEtaXK}{\ensuremath{\Bp \to \eta_X \Kp}\xspace}

\def\Beta#1#2#3#4K{\ensuremath{\Bp \to \eta(#1#2#3#4)\Kp}\xspace}
\def\Bfone#1#2#3#4K{\ensuremath{\Bpm \to \f_1(#1#2#3#4)\Kpm}\xspace}
\def\Bx#1#2#3#4K{\ensuremath{\Bpm \to \x#1#2#3#4\Kpm}\xspace}
\def\Bphi#1#2#3#4K{\ensuremath{\Bpm \to \phi(#1#2#3#4)\Kpm}\xspace}

\def\etapipi{\ensuremath{\eta\pipi}\xspace}
\def\etapp{\ensuremath{\eta\pi\pi}\xspace}

\def\kkst{\ensuremath{\kaon\Kstarb}\xspace}

\def\jp#1#2 {\ensuremath{J^P=#1^{#2} }\xspace}

\newcommand{\fisherL}{\ensuremath{\mathcal{F}_L}\xspace}

\def\hlct        {\ensuremath{\mathcal H}\xspace}

\def\win1 {\ensuremath{(1.2-1.4)\gev}\xspace}

\newcommand{\xv}{\mbox{\boldmath$x$}}
\newcommand{\betav}{\mbox{\boldmath$\zeta$}}

\newcommand{\BABARPubYear}     {07}
\newcommand{\BABARPubNumber}  {043}

\newcommand{\SLACPubNumber} {12840}

\begin{document}

\begin{flushleft}
\babar-PUB-\BABARPubYear/\BABARPubNumber\\
SLAC-PUB-\SLACPubNumber
\\[10mm]
\end{flushleft}

\title{
\large \bfseries \boldmath
Study of $B$ Meson Decays with Excited $\eta$ and $\eta^\prime$ Mesons
}

%
\author{B.~Aubert}
\author{M.~Bona}
\author{D.~Boutigny}
\author{Y.~Karyotakis}
\author{J.~P.~Lees}
\author{V.~Poireau}
\author{X.~Prudent}
\author{V.~Tisserand}
\author{A.~Zghiche}
\affiliation{Laboratoire de Physique des Particules, IN2P3/CNRS et Universit\'e de Savoie, F-74941 Annecy-Le-Vieux, France }
\author{J.~Garra~Tico}
\author{E.~Grauges}
\affiliation{Universitat de Barcelona, Facultat de Fisica, Departament ECM, E-08028 Barcelona, Spain }
% ...
\collaboration{The \babar\ Collaboration}
\noaffiliation

\date{\today}

\begin{abstract}
Using $383$~million \BB pairs from the $\babar$ data sample,
we report results for branching fractions of six charged 
$B$-meson decay modes, where a charged kaon recoils against 
a charmless resonance decaying to $K\Kbar^*$ or $\eta\pi\pi$ 
final states with mass in the range $(1.2-1.8)\gevcc$.
We observe a significant enhancement at the low \kkst invariant 
mass which is interpreted as $\Bp\to\et1475\Kp$, 
find evidence for the decay $\Bp\to\et1295\Kp$, 
and place upper limits on the decays $B^+\to\eta(1405)K^+$, 
$B^+\to f_1(1285)K^+$, $B^+\to f_1(1420)K^+$, and $B^+\to\phi(1680)K^+$.
\end{abstract}

\pacs{13.25.Hw, 14.40.Cs, 12.39.St, 12.39.Mk}

\maketitle

Charmless hadronic $B$-meson decays have been of particular 
interest due to their sensitivity to weak interaction dynamics. 
The first observed gluonic-penguin--dominated decays, such as 
$B\to\eta^\prime K$
and $B\to\pi\kaon$~\cite{Behrens:1998dn},
allowed the study of $C\!P$ violation in these decays with 
potential sensitivity to new physics~\cite{Aubert:2006wv,Aubert:2007mj}.
The relatively large $B\to\eta^\prime K$ decay rate was also a topic of debate.
However, little is known about the $B$ meson decays to excited 
states of the $\eta$ and $\eta^\prime$ mesons. There are three candidates for 
the first excited states $\eta(1295)$, $\eta(1405)$, and 
$\eta(1475)$~\cite{bib:Yao2006}, and there is a possibility that they
might include a gluonium admixture~\cite{bib:glue}.
This part of the pseudoscalar meson spectrum remains 
uncertain after a few decades of studies
\cite{bib:glue,Bai:1990hs,Augustin:1990ki,Foster:1969di,Adams:2001sk,Nichitiu:2002cj,Ahohe:2005ug}.
A search for $B$-meson decays to these pseudoscalar
states is the focus of this Letter. 

The $\eta$ and $\eta^\prime$ candidates and their excited counterparts,
which we call generically
$\etaX$ in this paper, have the quantum numbers $J^P=0^-$ and 
decay strongly to at least three pseudoscalar mesons. Thus we look for the 
$\etaX\to K\Kbar\pi$ and $\eta\pi\pi$ final states.
In the former case, the resonant structure $\kkst+\Kbar K^*$ is of particular 
interest and we refer to it as $\kkst$. Previously, the $\Kstar\Kp\Km$ final state
has been studied by $\babar$ inclusively \cite{Aubert:2006aw}.
The $J^P=1^+$ mesons $f_1(1285)$ and $f_1(1420)$ and $J^P=1^-$ meson $\phi(1680)$ 
also appear in the mass range $(1.2-1.8)\gevcc$ in these final states.
These resonances are considered in our search for the decays $B^+\to\etaX K^+$
and referred to by the generic nomenclature \etaX as well.
Hermitian conjugation is implied throughout this paper unless stated otherwise.

The $B\to\etaX K$ decay mechanism is expected to be dominated 
by the $b\to s$  gluonic-loop penguin diagram, similar to the 
$B\to\eta^\prime K$ decay. The expected branching fractions differ 
significantly depending on the \etaX state~\cite{bib:Yao2006},
following a pattern that early naive factorization models were unable to 
predict~\cite{Bauer:1986bm}.
The first attempt at unraveling the pattern in the branching fractions
of $B$-meson decays with $\eta$ and \etapr \cite{bib:Lip91}
suggested including the interference within the quark flavor octet among other
possible scenarios,
but the predictions did not match the experimental data.
More recent calculations find a larger predicted rate for $B\to\eta^\prime 
K$, in agreement with data, with inclusion of higher-order corrections 
\cite{bib:Ben2003} or ``charming-penguin'' contributions \cite{Williamson:2006hb}; large 
theoretical uncertainties persist, partly due to insufficient experimental data.
An admixture of a bound two-gluon state, gluonium,
in \etaX could also explain
the enhancement of the branching fractions.

Although the \et1295, \et1405, and \et1475 states are considered
well-established~\cite{bib:Yao2006}, their nature is still unknown.
Partial wave analyses of the $K\Kbar\pi$ and $\eta\pi\pi$ spectra
from past experiments, such as studies in Refs.~\cite{Bai:1990hs,
Augustin:1990ki,Adams:2001sk,Nichitiu:2002cj},
conclude that the meson spectrum in the $(1.2-1.8)\gevcc$ range
is described by a linear combination of the resonant states and
a nonresonant phase-space contribution. The analyses in
Refs.~\cite{Bai:1990hs,Augustin:1990ki} found that mass spectrum
description without interference between the resonant and nonresonant
contributions is preferred. Therefore, in our analysis we adopt the
model of three spin-zero resonances \et1295, \et1405, \et1475,
three spin-one resonances  $f_1(1285)$, $f_1(1420)$, $\phi(1680)$,
and a phase-space nonresonant contribution without interference
with the above states. Only four resonances are considered in each
final state, $K\Kbar^*$ or $\eta\pi\pi$, according to their dominant
decay modes as discussed below.

We use a sample of $(383\pm 4)$ million $\FourS\to\BB$ events
collected with the $\babar$ detector~\cite{bib:Aub2002} at the PEP-II $e^+e^-$ 
asymmetric-energy storage rings with
the $e^+e^-$ center-of-mass energy $\sqrt{s} = 10.58$ GeV.
Momenta of charged particles are measured 
in a tracking system consisting of a silicon vertex tracker with five 
double-sided layers and a 40-layer drift chamber, both within the 1.5-T 
magnetic field of a solenoid. 
Identification of charged particles is provided 
by measurements of the energy loss in the tracking devices and by 
a ring-imaging Cherenkov detector. 
Photons are detected by a CsI(Tl) electromagnetic calorimeter.

We search for $\Bp\to\etaX\Kp$ where \etaX decays to \kkst and $\etapipi$.
We reconstruct $\kkst\to\KorKbar^0 K^\pm\pi^\mp$, 
$\KorKbar^0\to\KS\to\pipi$, and $\eta\to\gamma\gamma$.
Isospin symmetry implies that the final states $\Kz\Km\pip+\Kzb\Kp\pim$
and $\eta\pipi$ constitute two thirds of $\etaX\to\kkst$ and $\etaX\to\etapp$,
respectively.

We identify $B$ meson candidates using two kinematic variables:
$\mes = \sqrt{s/4-\mathbf{p}_B^2}$
and $\DeltaE = \sqrt{s}/2 - E_B$,
where $(E_B,\mathbf{p}_B)$ is the four-momentum of the $B$ candidate
in the \epem center of mass frame.
We require $m_{\rm{ES}}>5.25\gevcc$ and $|\Delta{E}|<0.1\gev$.
The requirements on the invariant masses are
$1.35 < m_{\kkst} < 1.8\gevcc$,
$1.2 < m_{\etapp} < 1.5\gevcc$,
$|m_{\pi\pi}-m_{K^0}|< 12\mevcc$, and
$510<m_{\gamma\gamma} < 570\mevcc$.
The $\etaX$ invariant mass range is chosen to include
the broad spectrum of states without extending it above the
charm background production threshold.

We require the photon energies be at least $100\mev$.
For the $\KS$ candidates, we require the cosine of the angle
between the flight direction from the interaction point
and the momentum direction to be greater than $0.995$,
and the measured proper decay time to be greater than $5$ times its
uncertainty.
In the $\etaX\to\kkst+\Kbar K^*\to K\Kbar\pi$ decay channel, we require
the $\Kbar\pi$ or $K\pi$ invariant mass to satisfy
$0.85 < m_{K\pi}<0.95\gevcc$ for either
$K^\pm\pi^\mp$ or $\KorKbar^0\pi^\mp$ combinations.

We use the angle $\theta_T$ between the $B$-candidate
thrust axis and that of the rest of the event, and a Fisher
discriminant \fisherL to reject the dominant $e^+e^-\to$ 
quark-antiquark background~\cite{Aubert:2004ih}. 
Both variables are calculated in the \epem center-of-mass frame.
The discriminant combines the
polar angles of the $B$-candidate momentum vector and its thrust
axis with respect to the beam axis, and two moments
of the energy flow around the
$B$-candidate thrust axis~\cite{Aubert:2004ih}.

We suppress the background from $B$-decays into states with $D$ or $c\bar{c}$
mesons by applying vetos on the invariant masses of their decay products.
The remaining background (less than $10\%$) comes from random combinations of tracks
from $B$ decays, and from $\Bp\to\kkst\Kp$.
When more than one candidate is reconstructed,
we select the one with the lowest combined $\chi^2$ of the charged-track vertex fit
and of the invariant mass of the $\KS$ or $\eta$ candidate relative to
the PDG values~\cite{bib:Yao2006}.

We define
the helicity angle $\theta_\hlct$ as the angle between the direction of
the $B$ meson and the normal vector to the \etaX three-body decay plane in
the \etaX rest frame.
The ideal distribution is
uniform, ${\cal H}^2$, or $(1-{\cal H}^2)$
for \etaX with $J^P=0^-$, $1^-$, or $1^+$, respectively, 
where ${\cal H}=\cos\theta_\hlct$.
The observed angular distribution can be parameterized
as a product of the ideal angular distribution
for a given spin and parity multiplied by
an empirical acceptance function parameterized as a polynomial
$P(|{\cal H}|)$.

We use an unbinned, extended maximum-likelihood fit
to extract the event yields $n_{j}$ and the parameters $\betav$ of the probability 
density functions (PDF) ${\cal P}_{j}$.
The index $j$ represents six event categories used in our data model:
the \bEtaXK signal 
(four categories in each of the two \etaX decay channels
as shown in Table~\ref{tab:results}),
combinatorial background (mostly $e^+e^-\to q\bar{q}$ production with a few percent
admixture of misreconstructed $B$-meson decays), and a possible background from 
$B\to\kkst\kaon$ (in the $\etaX\to\kkst$ channel)
or other $B$ backgrounds (in the $\etaX\to\etapp$ channel).
The likelihood ${\cal L}_i$ for each candidate $i$ is defined as
${\cal L}_i = \sum_{j}n_{j}{\cal P}_{j}(\xv_i,\betav)$,
where the PDF is formed from the observables
$\xv =\{\mes, \DeltaE, \fisherL, {\cal H}, m\}$.
Here $m$ is the invariant mass of the \etaX candidate.

\begingroup
\begin{figure}[t]
\centerline{
\setlength{\epsfxsize}{\linewidth}\leavevmode\epsfbox{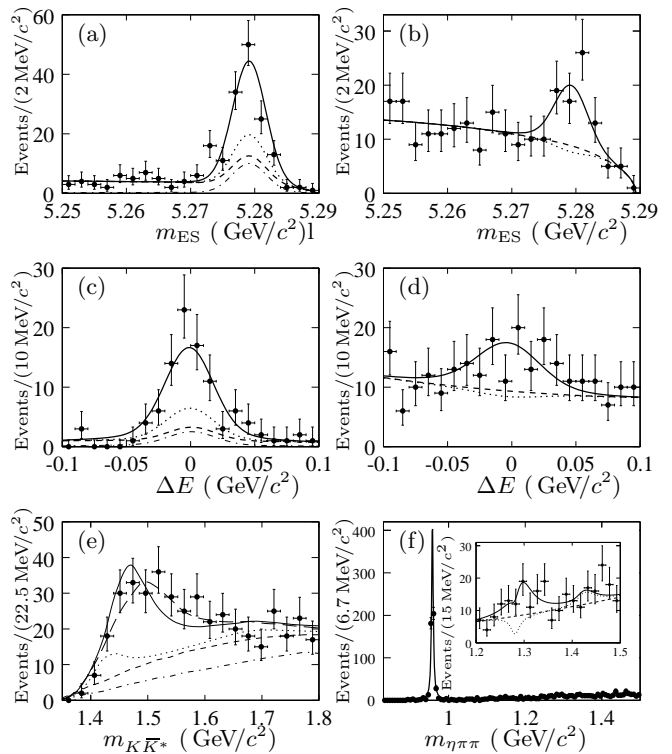}
}
\vspace{-0.3cm}
\caption{\label{fig:projection1} 
Projections for $\Bp\to\kkst\Kp$ (left column) and $\Bp\to\etapp\Kp$ (right column) of
(a,b)~\mes,
(c,d)~\DeltaE,
(e,f)~$m$
with a requirement applied on the signal-to-background probability ratio
calculated with all variables except the one being plotted.
The extended mass region in (f) includes the \etapr resonance
as a crosscheck. The nominal region is shown in the inset.
The solid (dashed) lines show the signal-plus-background
(background) PDF projections.
The dotted line shows the total PDF projection excluding
the $\et1475 K^+$ (left) or $\et1295 K^+$ (right) final states.
The dash-dotted lines indicate the nonresonant component.
The long-dashed line in (e) represents
the cross-check with the \et1475 resonance mass ($m_0$)
and width ($\Gamma$) parameters unconstrained, both resulting in
larger values.
}
\end{figure}
\endgroup
\begingroup
\begin{table*}[ht]
\caption{\label{tab:results}
Summary of results for the \bEtaXK process studied with 
six $B$-decay modes and eight decay channels with the signal
resonance and nonresonant model discussed in text, where 
$\etaX\to\kkst\to\KS\Kpm\pimp$ in the upper part and
$\etaX\to\etapipi$ in the lower part.
The mass $m_0$ and width $\Gamma$ of six $\eta_X$ states are quoted~\cite{bib:Yao2006} 
with errors in parentheses. The number of signal events $n_\mathtt{sig}$
with significance of the observed signal in parentheses, 
the product of the branching fractions $\BR$ and the corresponding daughter branching fractions,
the $B^+\to f_1(1285)K^+$ branching fraction, the corresponding $90\%$~C.L. upper limits,
and selection efficiencies $\epsilon$ obtained from MC simulation are shown.
The systematic uncertainties are quoted last.
}
\begin{center}
{
\begin{ruledtabular}
\setlength{\extrarowheight}{1.5pt}
\begin{tabular}{lcccc}
$\etaX\to\kkst$       & $\et1475$ &   $\ph1680$  &       $\et1405$ &   $\fone1420$  \\
$m_0/\Gamma$~\cite{bib:Yao2006}, \mev  	& $1476(4)/87(9)$ &	$1680(20)/150(50)$ & $1409.8(2.5)/51.1(3.4)$ & $1426.3(0.9)/54.9(2.6)$ \\
$n_\mathtt{sig}$  &	$155^{+21+11}_{-19~-6}~(7.5\sigma)$&	$17^{+6}_{-9}\pm7$&	$-12^{+8}_{-5}\pm1$&	$36^{+13}_{-14}\pm7$\\
~~~~$90\%$~C.L.   &		$<192$&	$<39$&	$<12$&		$<56$\\
$\BR(\bEtaXK)\,\BR(\etaX\to\kkst)$&	$(13.8^{+1.8+1.0}_{-1.7-0.6})\,10^{-6}$&	$(1.5^{+0.5+0.7}_{-0.8-0.6})\,10^{-6}$&	$(-1.2^{+0.9}_{-0.5}\pm0.1)\,10^{-6}$	&	$(2.7^{+0.9}_{-1.0}\pm0.5)\,10^{-6}$\\
~~~~$90\%$~C.L. &	$<17\times10^{-6}$&	$<3.4\times10^{-6}$&	$<1.2\times10^{-6}$&	$<4.1\times10^{-6}$\\
$\epsilon$ ($\%$) &	$8.8\pm0.1$&	$9.0\pm0.2$&	$8.4\pm0.3$&	$10.7\pm0.3$\\
\vspace{-0.3cm}&&&&\\
&&&&\\
\vspace{-0.3cm}&&&&\\
$\etaX\to\eta\pi\pi$	&	$\et1295$&	$\fone1285$&	$\et1405$ &	$\fone1420$\\
$m_0/\Gamma$~\cite{bib:Yao2006}, \mev  	&	$1294(4)/55(5)$&	$1281.8(0.6)/24.2(1.1)$&	&	\\
$n_\mathtt{sig}$&	$131^{+35}_{-33}\pm10~(3.5\sigma)$&	$-30^{+21}_{-19}\pm14$&	$-14^{+36}_{-33}\pm6$&	$49^{+35}_{-34}\pm11$\\
~~~~$90\%$~C.L.&		$<179$&	$<30$&		$<54$&	$<99$\\
$\BR(\bEtaXK)\,\BR(\etaX\to\etapp)$&	$(2.9^{+0.8}_{-0.7}\pm0.2)\,10^{-6}$&	$(-0.8^{+0.6}_{-0.5}\pm0.4)\,10^{-6}$	&	$(-0.3^{+0.9}_{-0.8}\pm0.1)\,10^{-6}$&	$(1.4\pm1.0\pm0.3)\,10^{-6}$\\
~~~~$90\%$~C.L. &	$<4.0\times10^{-6}$&	$<0.8\times10^{-6}$&	$<1.3\times10^{-6}$&	$<2.9\times10^{-6}$\\
$\BR(B\to f_1(1285)K^+)$	&	---&	$(-1.5^{+1.1}_{-1.0}\pm1.2)\,10^{-6}$&	---&	---\\
~~~~$90\%$~C.L.&	--- &	$<2.0\times10^{-6}$&	---&	---\\
$\epsilon$ ($\%$)&	$17.6\pm0.3$&	$14.1\pm0.9$&	$16.5\pm1.2$&	$13.5\pm0.6$\\
\end{tabular}
\end{ruledtabular}
}
\end{center}
\end{table*}
\endgroup

We use a relativistic spin-$J$ Breit--Wigner amplitude parameterization 
for the invariant mass of an \etaX resonance with the nominal mass and
width parameters quoted in Table~\ref{tab:results}. We model the decay
kinematics as $\etaX\to\kkst\to K\Kbar\pi$ and $\etaX\to\az(980)\pi\to\etapp$.
For the $\eta_X\to\kkst$ mode, the $\eta_X$ invariant mass parameterization
is corrected for phase space of the $B^+\to \kkst K^+$ decay and averaged
over the $\Kstarb\to\Kbar\pi$ invariant mass values.
We ignore the interference between the overlapping
resonances because it averages to zero for resonances with different 
quantum numbers or because these resonances
have different final states, such  as $\eta(1405)$ and $\eta(1475)$. 
The former decays mainly to $a_0(980)\pi$ (or direct $K\Kbar\pi$) 
and the latter mainly to $\kkst$~\cite{bib:Yao2006}. 
We also ignore the interference between the resonant 
and nonresonant decays based on indications from previous studies 
of $\etaX$ decays~\cite{Bai:1990hs,Augustin:1990ki}
and due to
potentially different three-body structure. 
This interference effect would only increase the significance
estimate because the hypothesis of zero yield is not affected
and the likelihood of the nominal fit could only improve.
The significance is defined as the square root of the change in 
$2\ln{\cal L}$ when the yield is constrained to zero in the 
likelihood ${\cal L}$.

The signal PDF for a given candidate $i$ is the product of 
the PDFs for each of the discriminating variables.
The combinatorial background PDF is the product of the 
PDFs for independent variables. The signal and background PDFs 
are illustrated in Fig.~\ref{fig:projection1}.
We use a sum of Gaussian functions
for the parameterization of the signal PDFs 
for \DeltaE, \mes, and \fisherL.
For the combinatorial background, we use polynomials,
except for \mes and \fisherL distributions, which are parameterized by an empirical
phase-space function and by Gaussian functions, respectively.
The nonresonant $B\to\kkst\kaon$ 
background is parameterized
the same as signal, except for the quantity $m$, 
which is described by a phase-space
function.

The PDF parameters ($\betav$) of the combinatorial background
are left free to vary in the fit, 
except for
the parameters that describe
\fisherL and the \mes endpoint, which are fixed to the values
extracted from the data sideband region ($\mes<5.27\gevcc$ or $|\DeltaE|>0.07\gev$).
The PDF parameters for other event categories are taken from Monte Carlo (MC)
simulation~\cite{Agostinelli:2002hh} and adjusted with
$B\to\Dbar\pi$
calibration
data samples.
We allow the yields to become negative as long as the total 
likelihood function remains positive in the allowed ranges of the observables.
We study the goodness-of-fit and validate the fit procedure
using MC simulation and generated samples.

In Table~\ref{tab:results} we present the results of the fit.
We observe a large charmless contribution in the
$\Bp\to(\kkst)\Kp$ decay
with
a significant enhancement at the low \kkst invariant 
mass, which is interpreted as $\et1475\to\kkst$ from the decay
$\Bp\to\et1475\Kp$.
We also see evidence for a nonzero \Beta1295K 
yield in the $\eta(1295)\to\etapp$
channel.
The significances
are more than $7.5$ and $3.5$ standard deviations, respectively,
including systematic uncertainties.
The significance of the $\Beta1295K$ yield is obtained 
in the fit when all yields are restricted to be positive, thus
reducing the significance from the nominal fit.
The significance is calculated within the model of resonant and
nonresonant signal contributions discussed above and 
in earlier work~\cite{bib:Yao2006, Bai:1990hs, Augustin:1990ki}.
We quote 90\% confidence level (C.L.)
upper limits, taken to be the values below which lies 90\% of
the total of the likelihood integral in the positive branching fraction
or yield region.

We repeat the fit by varying the fixed parameters 
in {$\betav$} within their uncertainties 
to obtain the associated systematic uncertainties.
The biases from the presence of fake combinations
or other imperfections in the signal PDF model are estimated 
with MC simulation.
Additional systematic uncertainties originate
from other potential $B$ backgrounds, 
which we estimate can contribute 
at most a few events to the signal component.
  As a cross-check, we repeat the fit with the particle identification
  on the recoil kaon reversed in order to enhance the $B^+\to\eta_X \pi^+$
  topology by more than a factor of ten compared to the nominal
  reconstruction, and find no evidence for such a decay.
The systematic uncertainties in selection efficiencies are dominated 
by those in particle identification, track finding, and \KS
and $\eta$ selection. 
Other systematic effects arise from event-selection criteria, 
and the estimation of the number of $B$ mesons.

The states \et1475, \ph1680, and \fone1420 are expected to decay into
the $K\Kbar\pi$ final state through \kkst~\cite{bib:Yao2006}.
We cross-check the $\kkst$ dominance by removing the $\Kbar\pi$
mass requirement and find consistent results.
With the present dataset we are unable to resolve intermediate states
in the \etapp modes,
such as 
$\rho^0(770)$ and $a_0^\pm(980)$ resonances.

In the projection plots in Fig.~\ref{fig:projection1},
for illustration purposes,
the signal fraction is enhanced with a requirement on the 
signal-to-background probability ratio, calculated with 
the plotted variable excluded.
The $m$ projection plot in Fig.~\ref{fig:projection1}~(e) implies a possible
difference of the signal resonance parameters from the assumed values. 
We repeat the fit with the $\et1475$ resonance parameters
$m_0$ and $\Gamma$ unconstrained while constraining other fit parameters 
to the values from the nominal fit.
We find the $m_0$ and $\Gamma$ central values to be larger, 
but still consistent with the nominal values within statistical uncertainties 
($1482\pm10\mev$ and $108\pm20\mev$, respectively).
We also repeat the fit with the $m$ range extended up to 2.5 GeV/$c^2$
and find good extrapolation of the fit results in the full range, 
apart from the narrow charm production contribution 
just above the 1.8 GeV/$c^2$ threshold.

In summary, we have measured product branching fractions
$\BR(\bEtaXK)\times\BR(\etaX\to\kkst, \etapp)$ for six $B$-decay 
modes that have not been studied previously, where $\etaX$ stands
for either $\eta(1295)$, $\eta(1405)$, $\eta(1475)$, $f_1(1285)$,
$f_1(1420)$, or $\phi(1680)$.
We observe a significant enhancement at the low \kkst invariant 
mass which is interpreted as $\Bp\to\et1475\Kp$ and
find evidence for the decay $\Bp\to\et1295\Kp$.
These decays could be used to either test weak dynamics in
the predominant $b\to s$ gluonic-loop penguin transition or 
study the $\eta_X$ composition, including potential gluonium
admixture.

We are grateful for the excellent luminosity and machine conditions
provided by our \pep2\ colleagues, 
and for the substantial dedicated effort from
the computing organizations that support \babar.
The collaborating institutions wish to thank 
SLAC for its support and kind hospitality. 
This work is supported by
DOE
and NSF (USA),
NSERC (Canada),
CEA and
CNRS-IN2P3
(France),
BMBF and DFG
(Germany),
INFN (Italy),
FOM (The Netherlands),
NFR (Norway),
MES (Russia),
MEC (Spain), and
STFC (United Kingdom). 
Individuals have received support from the
Marie Curie EIF (European Union) and
the A.~P.~Sloan Foundation.

\end{document}